# Lagrangian Trajectory Modeling of Lunar Dust Particles

John E. Lane
*ASRC Aerospace, Kennedy Space Center, FL 32899, USA*
John.E.Lane@ksc.nasa.gov

Philip T. Metzger, A.M.ASCE
*NASA/KSC Granular Mechanics and Surface Systems Laboratory, KSC, FL 32899, USA*
Philip.T.Metzger@nasa.gov

Christopher D. Immer
*ASRC Aerospace, Kennedy Space Center, FL 32899, USA*
Christopher.D.Immer@nasa.gov

Xiaoyi Li
*Analex Corp., Kennedy Space Center, FL 32899, USA*
Xiaoyi.Li@ksc.nasa.gov

**Abstract**

A mathematical model and software implementation developed to predict trajectories of single lunar dust particles acted on by a high velocity gas flow is discussed. The model uses output from a computation fluid dynamics (CFD) or direct simulation Monte Carlo (DSMC) simulation of a rocket nozzle hot gas jet. The gas density, velocity vector field, and temperature predicted by the CFD/DSMC simulations, provide the data necessary to compute the forces and accelerations acting on a single particle of regolith. All calculations of trajectory assume that the duration of particle flight is much shorter than the change in gas properties, i.e., the particle trajectory calculations take into account the spatial variation of the gas jet, but not the temporal variation. This is a reasonable first-order assumption. Final results are compared to photogrammetry derived estimates of dust angles form Apollo landing videos.

**Introduction**

Apollo landing videos shot from inside the right LEM window, provide a quantitative measure of the characteristics and dynamics of the ejecta spray of lunar regolith particles beneath the Lander during the final 10 [m] or so of descent. Photogrammetry analysis gives an estimate of the thickness of the dust layer and angle of trajectory (Immer, 2008). In addition, Apollo landing video analysis divulges valuable information on the regolith ejecta interactions with lunar surface topography. For example, dense dust streaks are seen to originate at the outer rims of craters within a critical radius of the Lander during descent.

The primary intent of this work was to develop a mathematical model and software implementation for the trajectory simulation of lunar dust particles acted on by gas jets originating from the nozzle of a lunar Lander, where the particle sizes typically range from 10 μm to 500 μm. The high temperature, supersonic jet of gas that is exhausted from a rocket engine can propel dust, soil, gravel, as well as small rocks to high velocities. The lunar vacuum allows ejected particles to travel great distances unimpeded, and in the case of smaller particles, escape velocities may be reached. The particle size distributions and kinetic energies of ejected particles can lead to damage to the landing spacecraft or to other hardware that has previously been deployed in the vicinity. Thus the primary motivation behind this work is to seek a better understanding for the purpose of modeling and predicting the behavior of regolith dust particle trajectories during powered rocket descent and ascent.

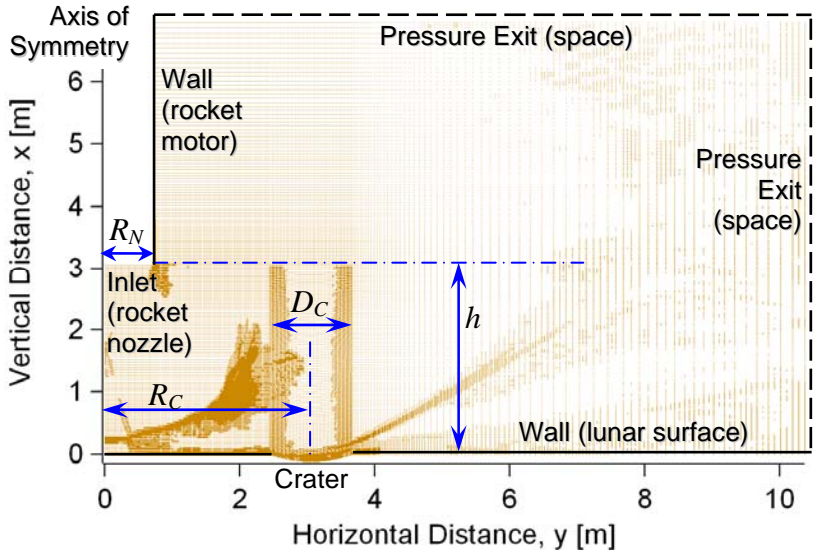

**Figure 1. 2D CFD boundary and grid point definition.**

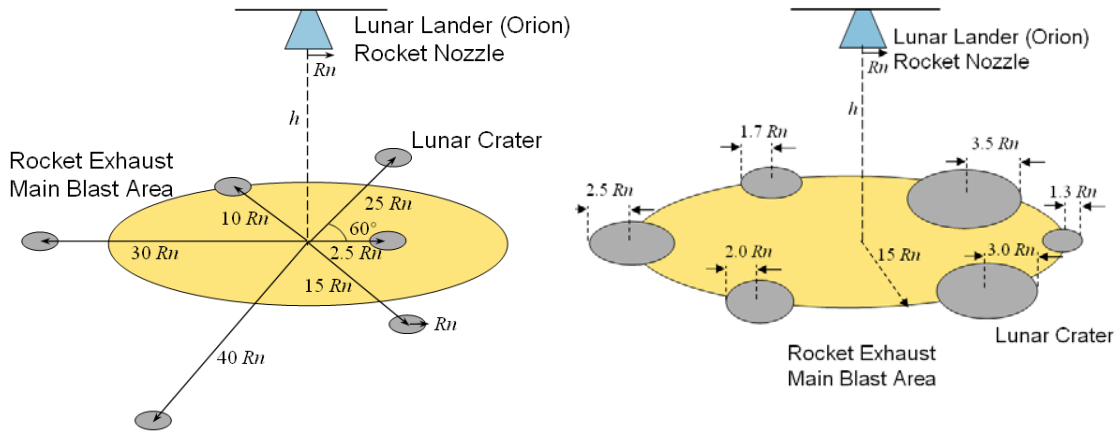

**Figure 2. 3D DSMC simulation geometry.**

## Computational Strategy

The trajectory model described in this paper requires output from a two- or three-dimensional fluid dynamics simulation of the rocket nozzle exhaust. Traditionally, this is achieved using a commercially available *computational fluid dynamics* (CFD) software package. Alternatively, *direct simulation Monte Carlo* (DSMC) software may be employed for expanding (rarefied) gas flow. DSMC simulators model fluid flows using a large number of discrete molecules in a Monte Carlo fashion to solve the Boltzmann gas equation. The worked described in this paper utilizes both CFD and DSMC simulations of an Apollo-like lunar Lander. In the case of the CFD simulations, 2D symmetry was exploited based on cylindrical symmetry where the problem is independent of azimuth angle, reducing complexity and computation time (see Figure 1). Full 3D geometries were processed by DSMC software, as depicted in Figure 2, where no symmetries were assumed (Lumpkin, 2007).

The CFD/DSMC output provides estimates of gas density $\rho(\mathbf{r})$, gas velocity $\mathbf{u}(\mathbf{r})$, and gas temperature $T(\mathbf{r})$ for a gridded volume described by $\mathbf{r}_i$ vector points in the bounded domain. These values are interpolated from the CFD/DSMC grid by finding four nearest grid neighbors in a volume around the *i*th trajectory point and applying an *N*-dimensional interpolation algorithm. Inputs which are initial conditions of the trajectory calculation include the particle diameter $D$ and initial position of the particle $\mathbf{r}_0 = (x_0, y_0, z_0)$ where the vertical direction $x$ is positive up and equal to zero at the surface. Typically, the particle starting position might be resting on the lunar surface at $x = 0$, so that $x_0 = D/2$.

**Mathematical Model and Trajectory Algorithm**

For a particle of diameter $D$ and mass $m$, the trajectory is due to three external forces: lunar *gravity*, jet gas *drag*, and *lift* caused by the gas flow. The sum of external forces is equal to the acceleration experienced by the particle, which can be estimated by a Taylor series expansion about time point $k$, resulting in a set of difference equations for position and velocity:

$$\mathbf{v}_k = \mathbf{v}_{k-1} + \mathbf{a}_{k-1}\Delta t \tag{1a}$$

$$\mathbf{r}_k = \mathbf{r}_{k-1} + \mathbf{v}_{k-1}\Delta t + \tfrac{1}{2}\mathbf{a}_{k-1}\Delta t^2 \tag{1b}$$

$$\mathbf{u}_k = \mathbf{u}(\mathbf{r}_k) \tag{1c}$$

$$\rho_k = \rho(\mathbf{r}_k) \tag{1d}$$

$$\begin{aligned}\mathbf{a}_k &= \frac{\pi C_D D^2}{8m}|\mathbf{u}_k - \mathbf{v}_k|\cdot(\mathbf{u}_k - \mathbf{v}_k)\rho_k + \mathbf{a}_{Lift} + g_L\hat{\mathbf{e}}_L \\ &= \frac{\pi C_D D^2}{8}\left(\frac{\pi D^3 \rho_L}{6}\right)^{-1}|\mathbf{u}_k - \mathbf{v}_k|\cdot(\mathbf{u}_k - \mathbf{v}_k)\rho_k + \mathbf{a}_{Lift} + g_L\hat{\mathbf{e}}_L \\ &= \frac{3 C_D \rho_k}{4D\rho_L}|\mathbf{u}_k - \mathbf{v}_k|\cdot(\mathbf{u}_k - \mathbf{v}_k) + \mathbf{a}_{Lift} + g_L\hat{\mathbf{e}}_L\end{aligned} \tag{1e}$$

$g_L$ is lunar gravity and $\rho_L$ is the lunar soil particle density (3100 [kg m$^{-3}$]); $\rho(\mathbf{r})$ is the gas density and $\mathbf{u}(\mathbf{r})$ is the gas velocity from the CFD/DSMC files. These values are computed by finding nearest neighbors in a volume around the point of interest and applying the spatial interpolation algorithm of Shepard (1968). $\mathbf{a}_{Lift}$ is the lift acceleration due to the vertical gradient of the horizontal component of gas flow (Shao, 2000):

$$\mathbf{a}_{Lift} \approx \frac{3C_L\rho_k}{2\rho_L}|\mathbf{u}_k - \mathbf{v}_k|_{horiz}\cdot\frac{\partial}{\partial x}|\mathbf{u}_k - \mathbf{v}_k|_{horiz} \tag{2}$$

where $C_L$ is the *coefficient of lift*. The direction of the lunar gravity unit vector is given by:

$$\hat{\mathbf{e}}_L \equiv \frac{-1}{\sqrt{(r_L+x)^2 + y^2 + z^2}} \begin{pmatrix} r_L + x \\ y \\ z \end{pmatrix} \quad (3)$$

where the lunar gravity constant $g_L$ is defined at the surface of a sphere of radius $r_L$ (moon radius). The coefficient of drag, $C_D$ is a function of the Reynolds number, $R$:

$$R_k = \frac{D\rho_k |\mathbf{u}_k - \mathbf{v}_k|}{A T_k^\beta} \quad (4)$$

where $T_k$ is the static temperature of the gas at the *k*th position of the particle. The empirical parameters in the denominator are: $A = 1.71575 \times 10^{-7}$ and $\beta = 0.78$. The coefficient of drag is computed from the following empirical formula:

$$C_D = \begin{cases} 24.0 R^{-1} & R < 2 \\ 18.5 R^{-0.6} & 2 \leq R < 500 \\ 0.44 & R \geq 500 \end{cases} \quad (5)$$

The initial conditions are:

$$\mathbf{v}_0 = 0, \quad \mathbf{r}_0 = \begin{pmatrix} x_0 \\ y_0 \\ z_0 \end{pmatrix}, \quad \mathbf{u}_0 = \mathbf{u}(\mathbf{r}_0), \quad \rho_0 = \rho(\mathbf{r}_0) \quad (6a)$$

$$\mathbf{a}_0 = \frac{3 C_D \rho_0}{4 D \rho_L} |\mathbf{u}_0| \cdot \mathbf{u}_0 + \frac{3 C_L \rho_k}{2 \rho_L} |\mathbf{u}_0|_{horiz} \cdot \frac{\partial}{\partial x} |\mathbf{u}_0|_{horiz} + g_L \hat{\mathbf{e}}_L \quad (6b)$$

**Simulation Experiments**

Numerous gas jet simulations involving variations of rocket nozzle height above ground, nozzle angles, and number of craters at various distances and diameters, were run using DSMC software. Some of these variations are depicted in the schematic of Figure 2. The DSMC output then became the *particle trajectory model* (PTM) input, where PTM is a FORTRAN code implementation of Eq. (1) through (6). These results were reported in detail by Metzger (2007). For the purpose of this present work, only CFD output is used as inputs to PTM. The four CFD test cases used in the trajectory simulation experiments are summarized in Tables 1. Figures 3, 4, and 5 display the CFD Case 2 output array values needed for input to PTM, i.e., gas *density*, gas *velocity*, and gas *temperature*.

| | Distance from Nozzle Center Line to Crater Center, $R_C$ | |
|---|---|---|
| Height of Nozzle above Surface, $h$ | 5 $R_N$ | 15 $R_N$ |
| 2.5 $R_N$ | Case C2 | --- |
| 5 $R_N$ | Case C7 | Case C3 |
| 10 $R_N$ | Case C1 | --- |

**Table 1. CFD Cases used as Inputs to PTM.**

| $D$ [μm] Particle diameter | $x$ [m] Initial vertical pt above ground | $y$ [m] Initial radial pt from nozzle | $z$ [m] Not used in 2D-CFD case | $\Delta t$ [μs] Time step | $N_L$ Number of iterations | $C_L$ Coefficient of lift |
|---|---|---|---|---|---|---|
| 50 | 0.003 | 3.7762 | 0 | 1000 | 500 | 20 |

**Table 2. Example PTM input parameters.**

| $D$ [μm] | 10 | 25 | 50 | 100 | 250 | 500 |
|---|---|---|---|---|---|---|
| $\Delta t$ [μs] | 200 | 500 | 1000 | 2000 | 5000 | 10000 |

**Table 3. PTM time step versus particle diameter.**

Figure 6 shows the PTM output for six particle sizes given by Table 3, spanning the range of lunar dust particle sizes. The initial starting coordinate in Table 2 is used in all six runs and corresponds to a point 3 [mm] above the outer lip of the crater. The reason for choosing a point slightly above the ground will be discussed in the following section. Similar plots are generated for the other cases shown in Table 1.

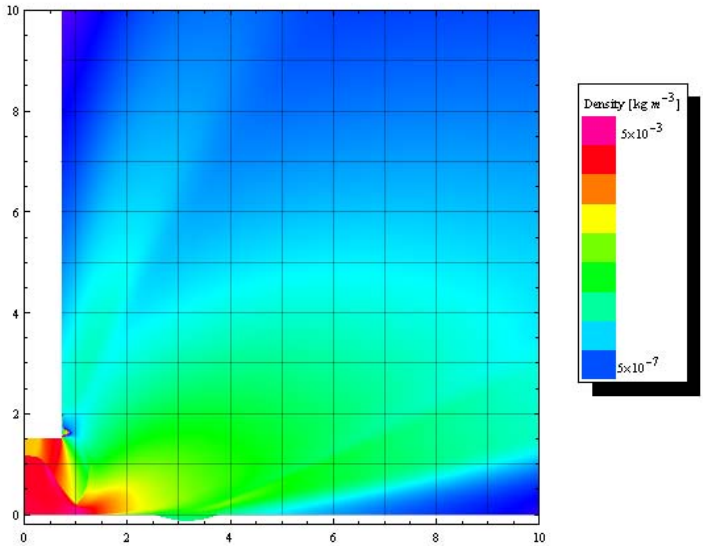

**Figure 3. Example gas density plot corresponding to CFD Case C2. Horizontal and vertical axes are marked in meters.**

In order to compare results from the four CFD cases studied above, the velocity vector corresponding to the trajectories of each of the six particle sizes is calculated from the trajectory data. The horizontal right boundary at 10 [m] is a convenient point to evaluate velocity. Figure 8 is a particle velocity magnitude (speed) plot of three CFD cases, corresponding to three rocket nozzle heights, $h$, above the ground compared with DSMC data from Metzger (2007). Figure 9 is the particle velocity angles relative to the ground for the three CFD cases.

PTM plots for six particle sizes are shown in Figure 7, corresponding to Case C3. In Case C3, the crater is located at $15R_N \approx 30$ [ft]. Initializing the starting point of the particles according to Table 2 results in velocities with lower angles and speeds. This result is a consequence of the turbulence generated above the crater, slowing down the particle with a downward component of velocity.

## Discussion and Conclusions

The particle velocities converge nicely for sizes in the range of 200 to 500 [μm], as shown by Figure 8. For smaller particle sizes, there is a trend that shows higher velocities for larger nozzle height above the ground for the three cases considered, i.e., $h$ = 5, 10, and 20 [ft]. The velocity magnitudes in this region are significantly larger than those shown by Metzger (2007). This may be attributed to the particle starting point: in the present case, particles were stated from the crater outer rim,

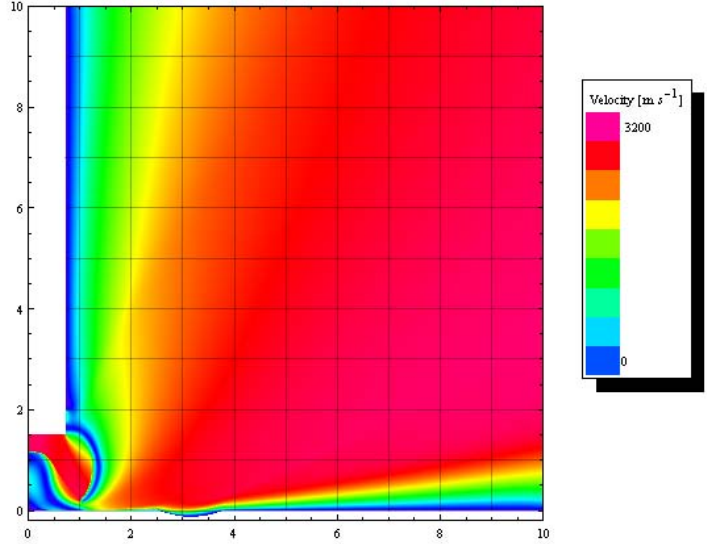

**Figure 4. Example gas velocity plot corresponding to CFD Case C2. Horizontal and vertical axes are marked in meters.**

whereas in Metzger's data, the starting point was typically not on a crater rim.

The CFD trajectory angle results shown in Figure 9 agree well with the photogrammetry results of Immer (2008). However, these results are not in good agreement with Metzger (2007), except for a narrow region around particle sizes of 200 [μm]. The source of this disagreement again may be due to differences in DSMC versus CFD simulations. It should also be noted that the *coefficient of lift* used in the current work is set to $C_L = 20$, whereas the previous work by Metzger, $C_L = 500$.

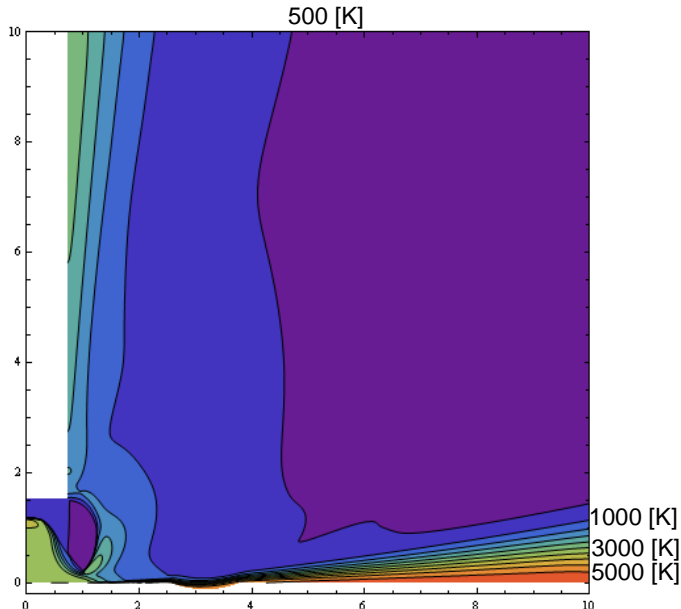

**Figure 5. Example temperature contour plot corresponding to Case C2. Horizontal and vertical axes are marked in meters.**

In the absence of a vertical component of drag, which is a reasonable assumption very near the ground (if the ground is smooth), the lift force must overcome the weight of the particle in order to lift the particle into the gas stream where horizontal drag takes over and dominates particle trajectory:

$$a_{Lift} = C_L \frac{3\rho_k}{2\rho_L} u'_k \frac{\partial u_k}{\partial x} > g_L \quad (7)$$

where $g_L$ is lunar gravity (1.622 [m s$^{-2}$]), $\rho_L$ is the lunar soil particle density (3100 [kg m$^{-3}$]), $u'_k$ is the particle speed relative to

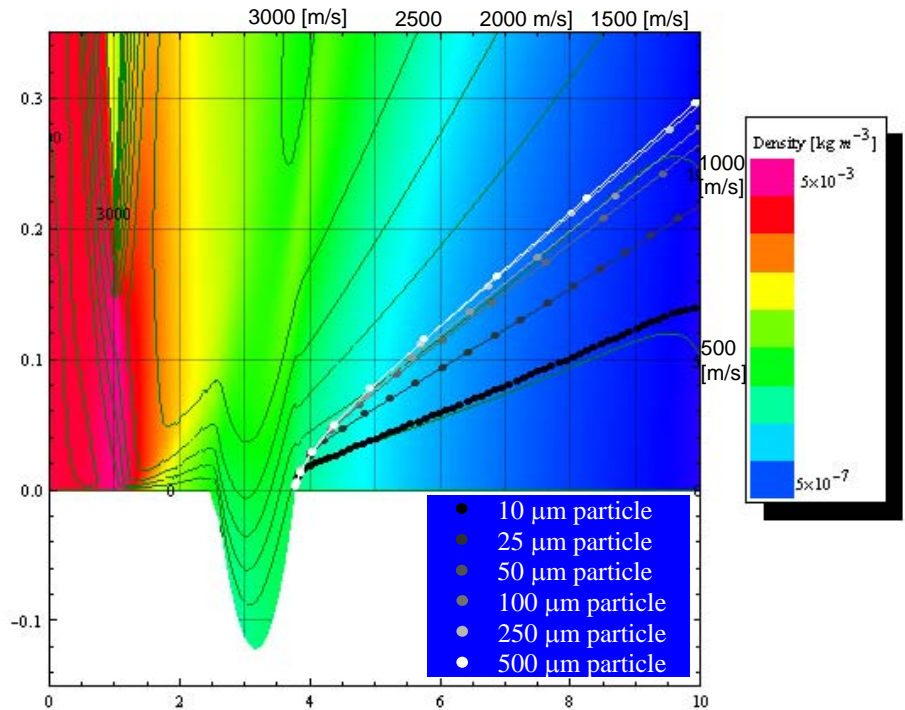
**Figure 6. PTM output for CFD Case 2 and initial values from Table 2. Horizontal and vertical axes are marked in meters.**

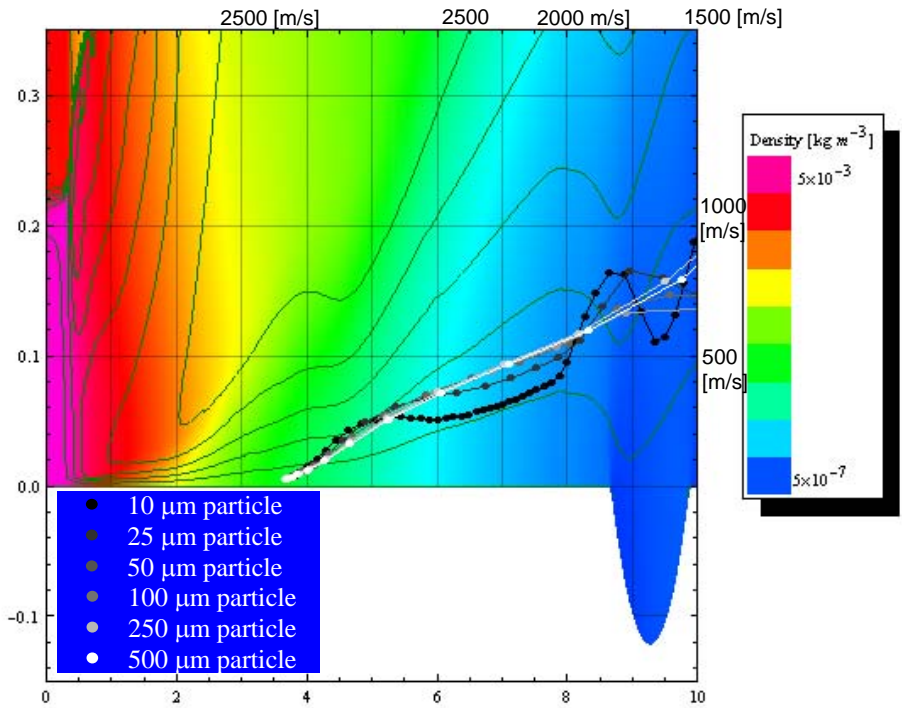
**Figure 7. PTM output for CFD Case 3 and initial values from Table 2. Horizontal and vertical axes are marked in meters.**

gas speed at time step $k$, and $\partial u'_k / \partial x$ is the gradient of $u'_k$ in the vertical direction. For example, using CFD Case 3 with the initial particle coordinates given by Table 2, $u'_0 \approx 110 \, [m \, s^{-1}]$ for a $D = 50$ [μm] diameter particle, $\partial u'_k / \partial x \approx 10/D \, [s^{-1}]$, and $\rho_0 \approx 2 \times 10^{-5} [kg \, m^{-3}]$. Evaluating Eq. (7) with these values with $C_L = 20$, $a_{Lift} \approx 4.3 \, [m \, s^{-2}] > g_L$. Therefore, the particle experiences lift.

As a counter example, using the same initial conditions as above but let the vertical initial coordinate start with the particle resting on the ground, $D/2 = 25 \, [\mu m]$ where $D$ is the particle diameter: $u'_0 \approx 10 \, [m \, s^{-1}]$, again $\partial u'_k / \partial x \approx 10/D \, [s^{-1}]$, as well as $\rho_0 \approx 2 \times 10^{-5} [kg \, m^{-3}]$. Evaluating Eq. (7) with these values with $C_L = 20$, $a_{Lift} \approx 0.4 \, [m \, s^{-2}] < g_L$. In this case there is no lift. Note that using a larger value of $C_L = 500$ (Metzger, 2007), lift is achieved in this example.

The forces experienced by the particle are primarily due to drag, i.e., the particle will tend to follow the gas jet stream. However, particles on a perfectly smooth ground experience very little drag because of the *no slip* boundary condition which results in zero gas velocity at the ground boundary. Under these conditions, a lift force can artificially be generated in the trajectory simulation by using a very large value of lift coefficient (Metzger, 2007). More work will need to be done to better understand the conditions of the gas at the boundary. A hint of what might really be going on near the surface is observed by the disruption of flow caused by craters. On a smaller scale, any variations of the lunar surface are likely leading to boundary conditions that account for dust particle lift, beyond the simple model of aerodynamic lift presented above. Future work might also attempt to model the collision of particles leading to something like a "chain-reaction" causing particles to pop up into the gas jet streams from the surface. Once in the gas flow, they are easily dragged into a trajectory and ejected at the approximate 3 degree angle observed in the Apollo videos.

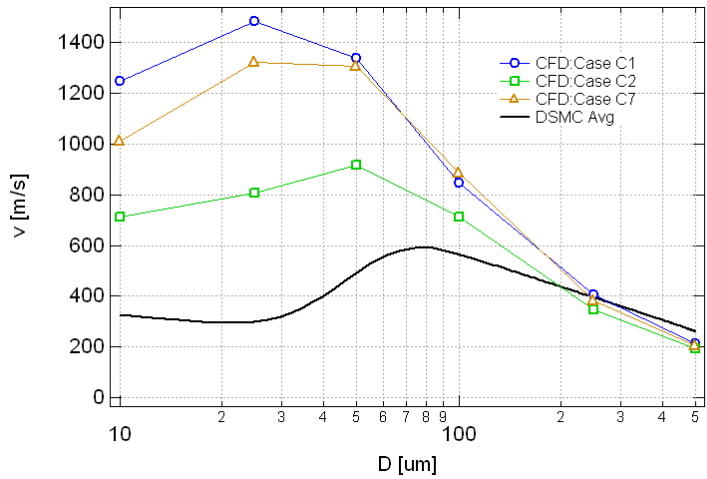

**Figure 8. Particle speeds exiting the CFD boundary for various particle sizes and CFD Cases of Table 1. The solid black line is averaged data from Metzger (2007).**

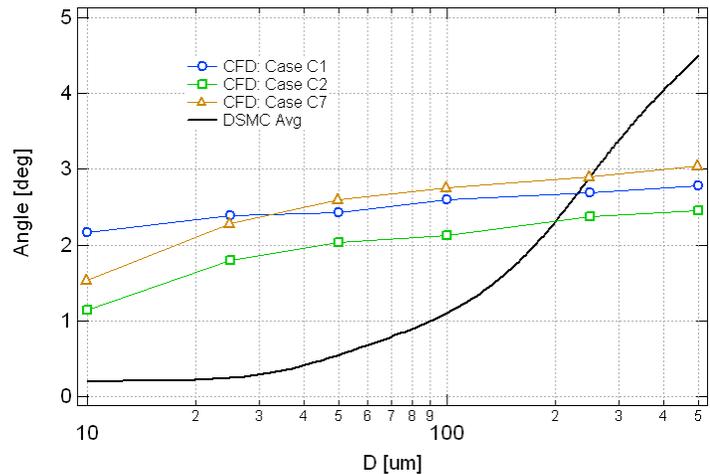

**Figure 9. Particle trajectory angles relative to ground corresponding to Figure 7. Again, the solid black line is averaged data from Metzger (2007).**